\documentclass{PoS}

\newcommand{\reseteqnum}{\setcounter{equation}{0}}

\newcommand{\ovl}[1]{\overline{#1}}

\newcommand{\eqn}[1]{(\ref{#1})}

\newcommand{\bpsi}{{\overline{\psi}}}

\newcommand{\pslash}{p\kern-1ex /}
\newcommand{\Dslash}{{\cal D}\kern-1.5ex /}

\title{Non-perturbative renormalization of quark mass in $N_f=2+1$ QCD
with the Schr\"odinger functional scheme}

\ShortTitle{Non-perturbative renormalization of quark mass in $N_f=2+1$
QCD with SF scheme}

\author{\speaker{Yusuke Taniguchi}%
\ \ for PACS-CS Collaboration
\\
Institute of Physics, University of Tsukuba,
Tsukuba, Ibaraki, 305-8571, Japan \\
        E-mail: \email{tanigchi@het.ph.tsukuba.ac.jp}}

\abstract{
 We present an evaluation of the quark mass renormalization factor for
 $N_f=2+1$ QCD. 
 The Schr\"odinger functional scheme is employed as the intermediate
 scheme to carry out non-perturbative running from the low energy
 to deep in the high energy perturbative region.
 The regularization independent step scaling function of the quark mass
 is obtained in the continuum limit. 
 Renormalization factors for the pseudo scalar density and the axial
 vector current are also evaluated for the same action and the bare
 couplings as two recent large scale $N_f=2+1$ simulations;
 previous work of the CP-PACS/JLQCD collaboration, which covered
 the up-down quark mass range heavier than $m_\pi\sim 500$~MeV
 and that of PACS-CS collaboration on the physical point using
 the reweighting technique.
}

\FullConference{The XXVIII International Symposium on Lattice Filed Theory\\
		 June 14-19,2010\\
		 Villasimius, Sardinia Italy}

\begin{document}

\reseteqnum
\section{Introduction}

The strong coupling constant and the quark masses constitute
fundamental parameters of the Standard Model.
It is an important task of lattice QCD to determine these
parameters using inputs at low energy scales.
In the course of evaluating these fundamental parameters we need the
process of renormalization in some scheme.
The ${\ovl{\rm MS}}$ scheme is one of the most popular schemes, and 
hence one would like to evaluate the running coupling constant and quark
masses through input of low energy quantities on the lattice and convert
them to the ${\ovl{\rm MS}}$ scheme.
A practical difficulty in this process is called the window problem
that the conversion should be performed at much higher energy
than the QCD scale.
At the same time the renormalization scale $\mu$ should be kept much
smaller than the lattice cut-off to reduce lattice artifacts:
$1/L\ll\Lambda_{\rm QCD}\ll\mu\ll{1}/{a}$.

The Schr\"odinger functional (SF) scheme
\cite{Luscher:1992an,Sint:2000vc} 
is designed to solve the window problem.
A unique renormalization scale is introduced through the box size $L$ in
the chiral limit and the scheme is mass independent.
A wide range of renormalization scales can be covered by the step
scaling function (SSF) technique.
This matches our goal to obtain the coupling constant and quark masses
in the $\ovl{\rm MS}$ scheme.
The SF scheme has been applied for evaluation of the $N_f=2+1$ QCD
coupling \cite{Aoki:2009tf}.

For the quark mass renormalization factor the SF scheme has been applied
for $N_f=2$ QCD \cite{Della Morte:2005kg}.
At low energy scales of $\mu\sim500$ MeV, where physical input is given,
we expect the strange quark contribution to be important in addition to
those of the up and down quarks.
Thus the aim of this report is to go one step further
and evaluate the quark mass renormalization factor in $N_f=2+1$ QCD
\footnote{This report is based on the published paper \cite{Aoki:2010wm}.}.
Our goal is to renormalize the bare light quark masses evaluated in two
recent large-scale $N_f=2+1$ lattice QCD simulations
\cite{Ishikawa:2007nn,Aoki:2008sm,Kuramashi} and derive the
renormalization group invariant (RGI) quark mass $M$.
Once the RGI mass $M$, which is scheme independent, is evaluated,
the conversion into the $\ovl{\rm MS}$ scheme can be carried out
perturbatively.

We therefore first derive the renormalization factor $Z_M(g_0)$, which
converts the bare PCAC mass at a bare coupling $g_0$ to the RGI mass.
Derivation of the renormalization factor proceeds in three steps
\cite{Della Morte:2005kg,Aoki:2010wm}
\begin{eqnarray}
Z_M(g_0,a/L_{\rm max})=
\frac{Z_A(g_0,a/L)}{Z_P(g_0,a/L_{\rm max})}
\frac{\ovl{m}(1/L_n)}{\ovl{m}(1/L_{\rm max})}
\frac{M}{\ovl{m}(1/L_n)}.
\label{eqn:Zm}
\end{eqnarray}
The first factor $Z_A/Z_P$ renormalizes the bare PCAC mass in the SF
scheme at an appropriately low energy scale $L_{\rm max}$.
The second factor ${\ovl{m}(1/L_n)}/{\ovl{m}(1/L_{\rm max})}$ represents
the running of the renormalized mass from $L_{\rm max}$ to a high energy
scale $L_n$, which is evaluated non-perturbatively in the SF scheme.
The last factor ${M}/{\ovl{m}(1/L_n)}$ represents the running from $L_n$
to infinitely high energy scale and is evaluated perturbatively for an
appropriately high energy scale $L_n$.

\reseteqnum
\section{Step scaling function}

We adopt the renormalization group improved Iwasaki gauge action and
the non-perturbatively ${O}(a)$ improved Wilson fermion action with the
clover term.
The Dirichlet boundary condition for the spatial gauge link is set to
\begin{eqnarray}
U_k(x)|_{x_0=0}=U_k(x)|_{x_0=T}=1
\end{eqnarray}
and the twisted periodic boundary condition of the fermion fields in the
three spatial directions is
\begin{eqnarray}
\psi(x+L\hat{k})=e^{i\theta}\psi(x),\quad
\bpsi(x+L\hat{k})=e^{-i\theta}\bpsi(x),\quad
\theta=0.5.
\end{eqnarray}

We prepare seven renormalized coupling values to cover weak to strong
coupling regions \cite{Aoki:2009tf}.
For each coupling we use three box sizes $L/a=4, 6, 8$ to take the
continuum limit.
At the three lattice sizes the values of $\beta$ and $\kappa$ were tuned
to reproduce the same renormalized coupling keeping the PCAC mass to
zero.
On the same parameters we evaluate the pseudo scalar density
renormalization factor ${Z_P(g_0,a/L)}$ and ${Z_P(g_0,a/(2L))}$ at two
box sizes.

The renormalization factor $Z_P$ is defined in terms of two-point
functions \cite{Della Morte:2005kg} of pseudo scalar density at bulk and
at boundary.
Taking the ratio of two	renormalization factors at two renormalization
scales $\mu=1/L$ and $1/(2L)$
we get the step scaling function (SSF) on the lattice
\begin{eqnarray}
\Sigma_P\left(u,\frac{a}{L}\right)
=\left.\frac{Z_P(g_0,a/(2L) )}{Z_P(g_0,a/L)}\right|_{\ovl{g}^2(L)=u,m=0},
\label{eqn:SSF}
\end{eqnarray}
where mass independent scheme in the massless limit is adopted.

We perform a perturbative improvement of the SSF before taking the
continuum limit.
For this purpose we need a perturbative evaluation of the lattice
artifact in the SSF
\begin{eqnarray}
\delta_{P}(u,a/L)=
\frac{\Sigma_{P}\left(u,{a}/{L}\right)-\sigma_{P}(u)}{\sigma_{P}(u)}.
\end{eqnarray}
Instead of calculating $\delta_P$ at one and two-loop level
perturbatively we calculate SSF's directly by Monte-Carlo simulations at
very weak coupling $\beta\ge10$.
The artifact is fitted to a polynomial form for each $a/L$,
\begin{eqnarray}
1+\delta_{P}(u,a/L)=1+d_{1}(a/L)u+d_{2}(a/L)u^2.
\end{eqnarray}
Since the quadratic fit provides a reasonable description of data
\cite{Aoki:2010wm} we opt to cancel the $O(a)$ contribution dividing out
the SSF by the quadratic fit.

\begin{figure}
 \begin{center}
  \includegraphics[width=5.0cm]{fig/sigmap.continuum3.eps}
  \includegraphics[width=5.0cm]{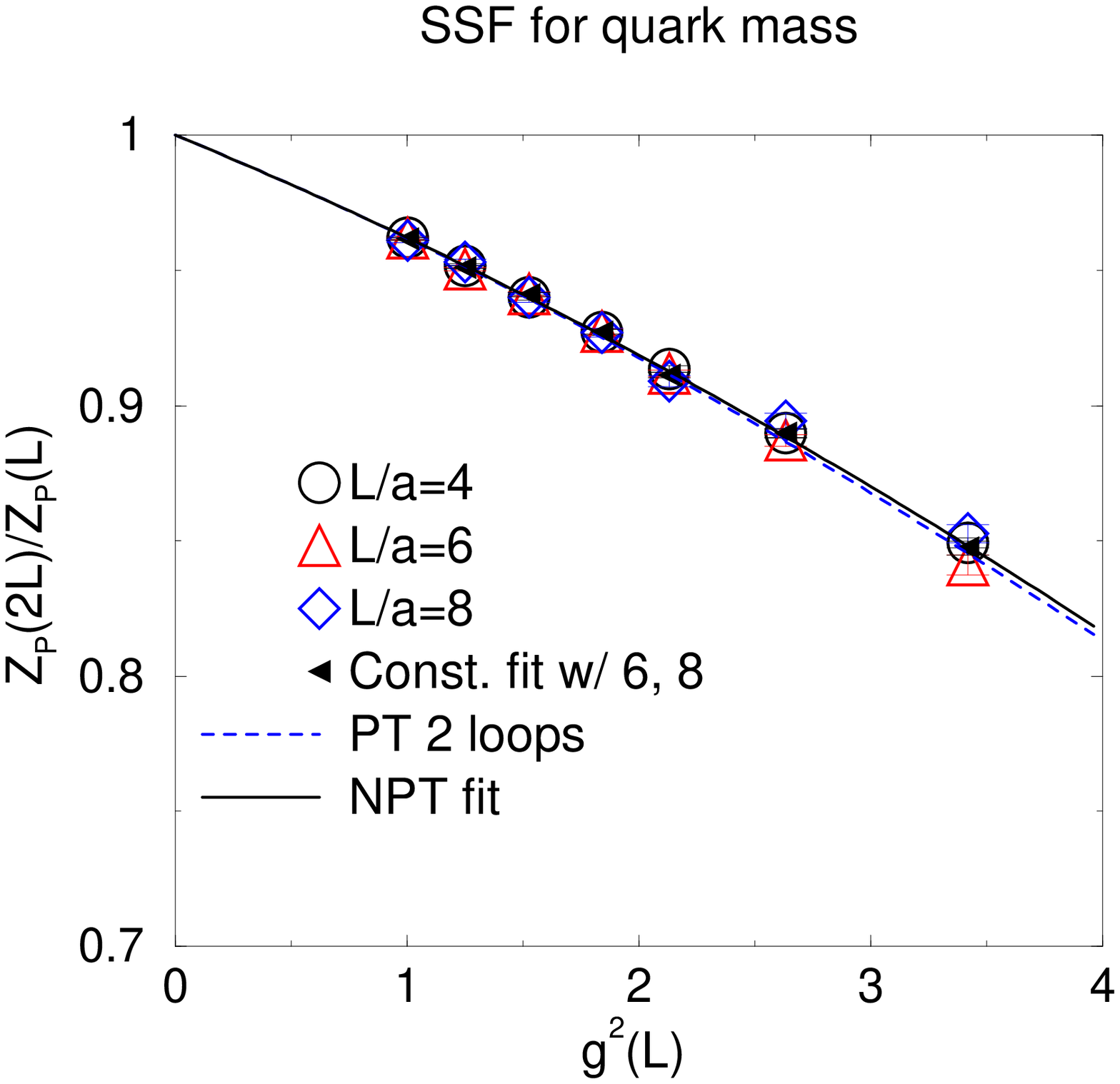}
  \caption{The SSF on the lattice with its continuum extrapolation at
  each renormalization scale (left) and the RG flow of the SSF (right).}
  \label{fig:SSF}
 \end{center}
\end{figure}

Scaling behavior of the improved SSF is plotted in the left panel of
Fig.~\ref{fig:SSF}.
Almost no scaling violation is found.
We performed three types of continuum extrapolation:
a constant extrapolation with the finest two (filled symbols) or all
three data points (open symbols), or a linear extrapolation with all
three data points (open circles), which are consistent with each other.
The constant fit was employed with the finest two data points to find
our continuum value.

The RG running of the continuum SSF is plotted in the right panel of
Fig.~\ref{fig:SSF}.
A polynomial fit of the continuum SSF to third order yields
\begin{eqnarray}
&&
\sigma_P(u)=1+p_0u+p_1u^2+p_2u^3,
\label{eqn:SSFP}
\\&&
p_1=-0.002826,\quad
p_2=0.000031
\end{eqnarray}
fixing the first coefficients $p_0$ to its perturbative value.
The fitting function is also plotted (solid line) together with the
two loops perturbative running (dashed line).

\begin{figure}
 \begin{center}
  \includegraphics[width=5.5cm]{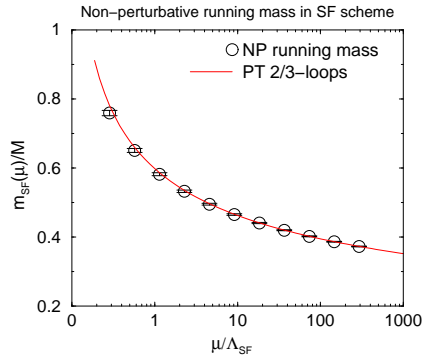}
  \caption{The non-perturbative running mass in the SF scheme.
  Solid line is a perturbative running with two and three loops RG
  function for the mass and the coupling.}
  \label{fig:running-mass}
 \end{center}
\end{figure}

Multiplying the SSF according to a sequence of couplings $u_i$, which
differ by a factor two in the renormalization scale, 
starting from $u_0=\ovl{g}^2(L_{\rm max})$ we get the non-perturbative
running of mass
\begin{eqnarray}
{\prod_{i=1}^n\sigma_P(u_i)}=\frac{\ovl{m}(1/L_n)}{\ovl{m}(1/L_{\rm max})}.
\end{eqnarray}
In Fig.~\ref{fig:running-mass} we plot the non-perturbative running mass
$\ovl{m}(1/L_n)/M$ in units of the RGI mass as a function of the scale
$\mu/\Lambda_{\rm SF}$, where $\mu=1/L_n$.

\reseteqnum
\section{$Z_P$ and $Z_A$ at low energy scale}

We evaluate the renormalization factor $Z_P$ at the same bare
coupling $\beta$, $1.83$, $1.90$ and $2.05$ adopted in the large scale
simulations \cite{Ishikawa:2007nn,Aoki:2008sm,Kuramashi}.
The reference scale $L_{\rm max}$ is given by the box size we used in
this evaluation.
The renormalized coupling $\ovl{g}^2(L_{\rm max})$ should
not exceed our maximal value $5.13$ for the coupling SSF significantly.
We use the box size of $L/a=4$ for $\beta=1.83$ and $1.90$ to define
$L_{\rm max}$ and $L/a=6$ for $\beta=2.05$.
We adopt the lattice spacing $a$ as an intermediate scale.

The remaining ingredient of the renormalization factor \eqn{eqn:Zm} at
low energy is $Z_A$ of the axial vector current.
We calculate the renormalization factor according to the procedure in
Ref.~\cite{Della Morte:2005rd} through the axial Ward-Takahashi
identity, which is applicable to small non-vanishing PCAC mass.
The box size $L$ for the axial current need not coincide with
$L_{\rm max}$ since $Z_A$ is scale independent in the continuum
limit.

\begin{figure}
 \begin{center}
  \includegraphics[width=5.0cm]{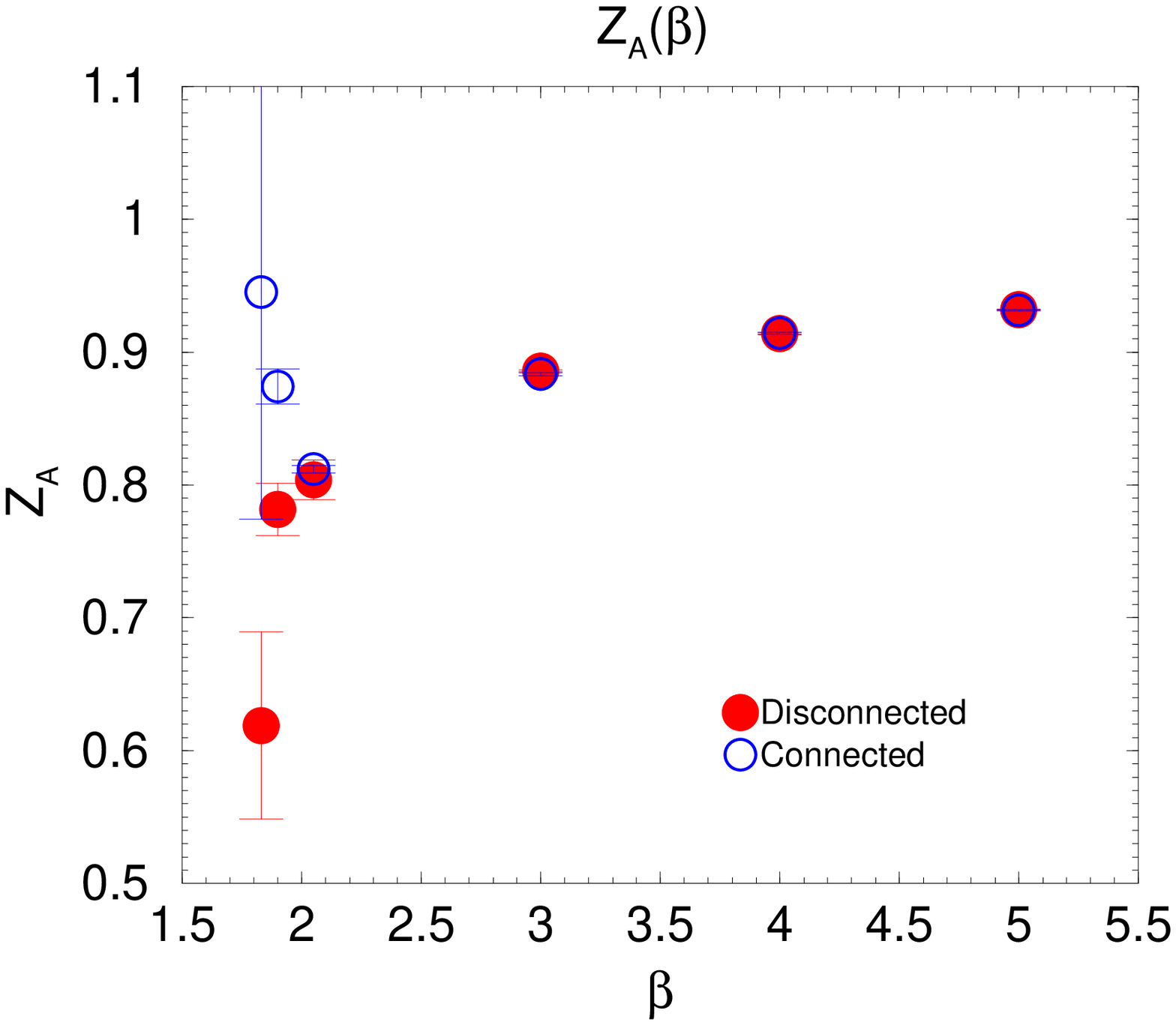}
  \includegraphics[width=5.0cm]{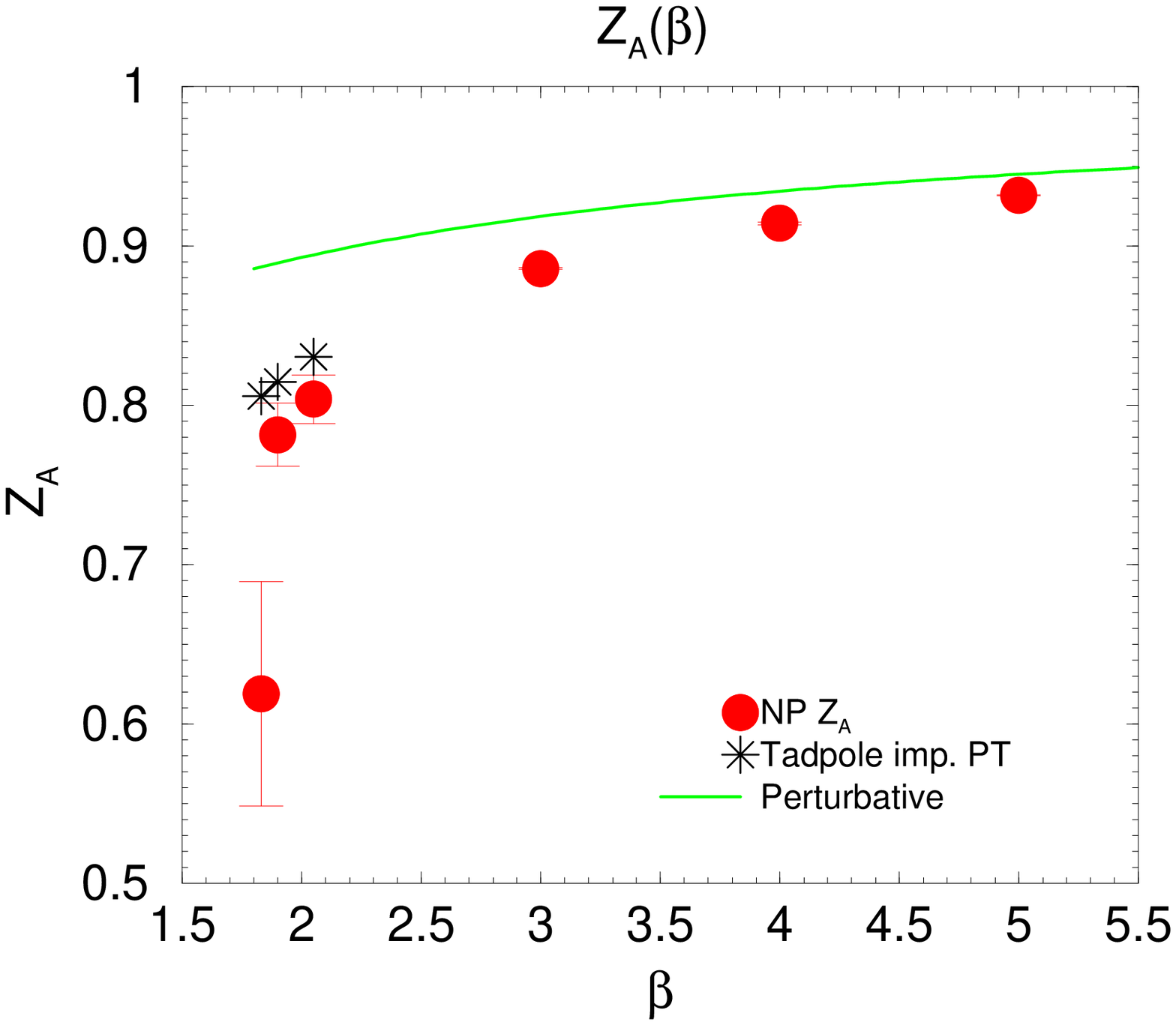}
  \caption{$\beta$ dependence of $Z_A(g_0)$ with and without
  disconnected diagrams (left) and comparison with perturbative results
  (right).}
  \label{fig:za-beta}
 \end{center}
\end{figure}

In this paper we adopted the following box size at each $\beta$ to
define $Z_A$:
$12^3\times30$, $\theta=0.5$ at $\beta=1.83$, $10^3\times24$, $\theta=0$
at $\beta=1.90$ and $12^3\times30$, $\theta=0.5$ for $\beta=2.05$.
The disconnected contribution \cite{Della Morte:2005rd} is included.
In the left panel of Fig.~\ref{fig:za-beta} we plot the values for $Z_A$
according to our 
definition (filled circles), together with those without the
disconnected contribution (open symbols).
Scattering of points starting around $\beta=1.95$ indicates that lattice
artifacts are increasingly large in our data for large lattice spacings.
In the right panel of Fig.~\ref{fig:za-beta} we compare our $Z_A(g_0)$
with perturbative behavior (solid line) and results from the tadpole
improved perturbation theory (star symbols).

\reseteqnum
\section{RGI mass renormalization factor}

We derive the renormalization factor $Z_M$ for the RGI mass, which is
intended to renormalize the bare PCAC masses obtained in the two large
scale simulations.
The three hadron masses, $m_\pi$, $m_K$, $m_\Omega$ were used in
Ref.~\cite{Aoki:2008sm,Kuramashi} to determine the light quark masses
and the lattice spacing.
Two choices $m_\pi$, $m_\rho$ $m_K$ or $m_\pi$, $m_\rho$ $m_\phi$ were
adopted in Ref.~\cite{Ishikawa:2007nn}.
The results for $Z_M$ are listed in Table \ref{tab:zm}.
Also listed are the renormalization factors
$Z_m^{\ovl{\rm MS}}(\beta,\mu=2\ {\rm GeV})$ in the $\ovl{\rm MS}$
scheme at a renormalization scale $\mu=2$ GeV.
We emphasize that the renormalization factor here is defined in terms
of the renormalization group functions for three flavors.
\begin{table}[htb]
\begin{center}
\begin{tabular}{|c|c|c|c|c|c|}
\hline
$\beta$
 & $Z_M$ & $Z_m^{\ovl{\rm MS}}$ ($K$)
 & PT(tad) $Z_m^{\ovl{\rm MS}}$ ($K$)
 & $Z_m^{\ovl{\rm MS}}$ ($\phi$)
 & PT(tad) $Z_m^{\ovl{\rm MS}}$ ($\phi$)\cr
\hline
$1.83$ \cite{Ishikawa:2007nn} & $1.33(15)$ & $1.04(12)$
 & $1.07161$
 & $1.04(12)$
 & $1.07019$\cr
$1.90$ \cite{Ishikawa:2007nn} & $1.693(46)$ & $1.315(35)$ 
 & $1.09973$
 & $1.315(35)$
 & $1.09955$\cr
$2.05$ \cite{Ishikawa:2007nn} & $1.862(41)$ & $1.417(29)$ 
 & $1.14487$
 & $1.416(29)$
 & $1.1446$\cr
\hline
$1.90$ \cite{Aoki:2008sm} & $1.693(46)$ & $1.344(36)$ & $1.11322$ & &\cr
$1.90$ \cite{Kuramashi} & $1.693(46)$ & $1.347(36)$ & $1.11322$ & &\cr
\hline
\end{tabular}
\caption{$Z_M$ for the RGI mass and $Z_m^{\ovl{\rm MS}}(2\ {\rm GeV})$
 in the $\ovl{\rm MS}$ scheme.
 ($K$) or ($\phi$) means which meson mass is used for physical scale
 input.
 Perturbative results with tadpole improvement are also listed.}
\label{tab:zm}
\end{center}
\end{table}

The error in the renormalization factor includes all the statistical
and systematic ones except for that from the choice of the reference
scale $L_{\rm max}$.
We tried a rough estimate of ${O}((a/L_{\rm max})^2)$ error and we find
a few percent effect at $\beta=2.05$, while the magnitude may increase
to a 10 \% level at lower values of $\beta$.
However, a firmer conclusion requires the step scaling function at the
couplings stronger than those explored in the present work.

\begin{figure}[htbp]
\begin{center}
 \includegraphics[width=5.5cm]{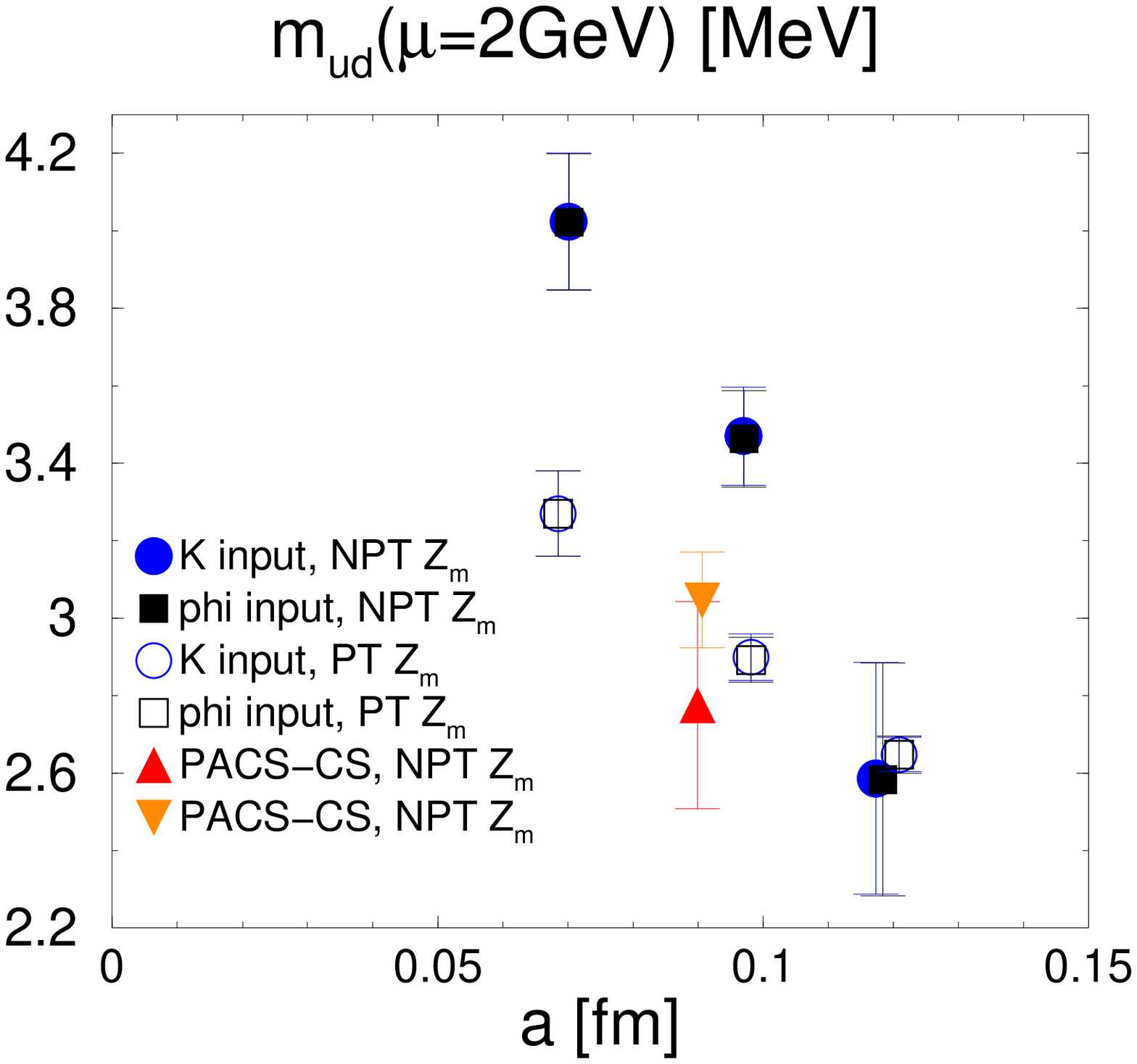}
 \includegraphics[width=5.5cm]{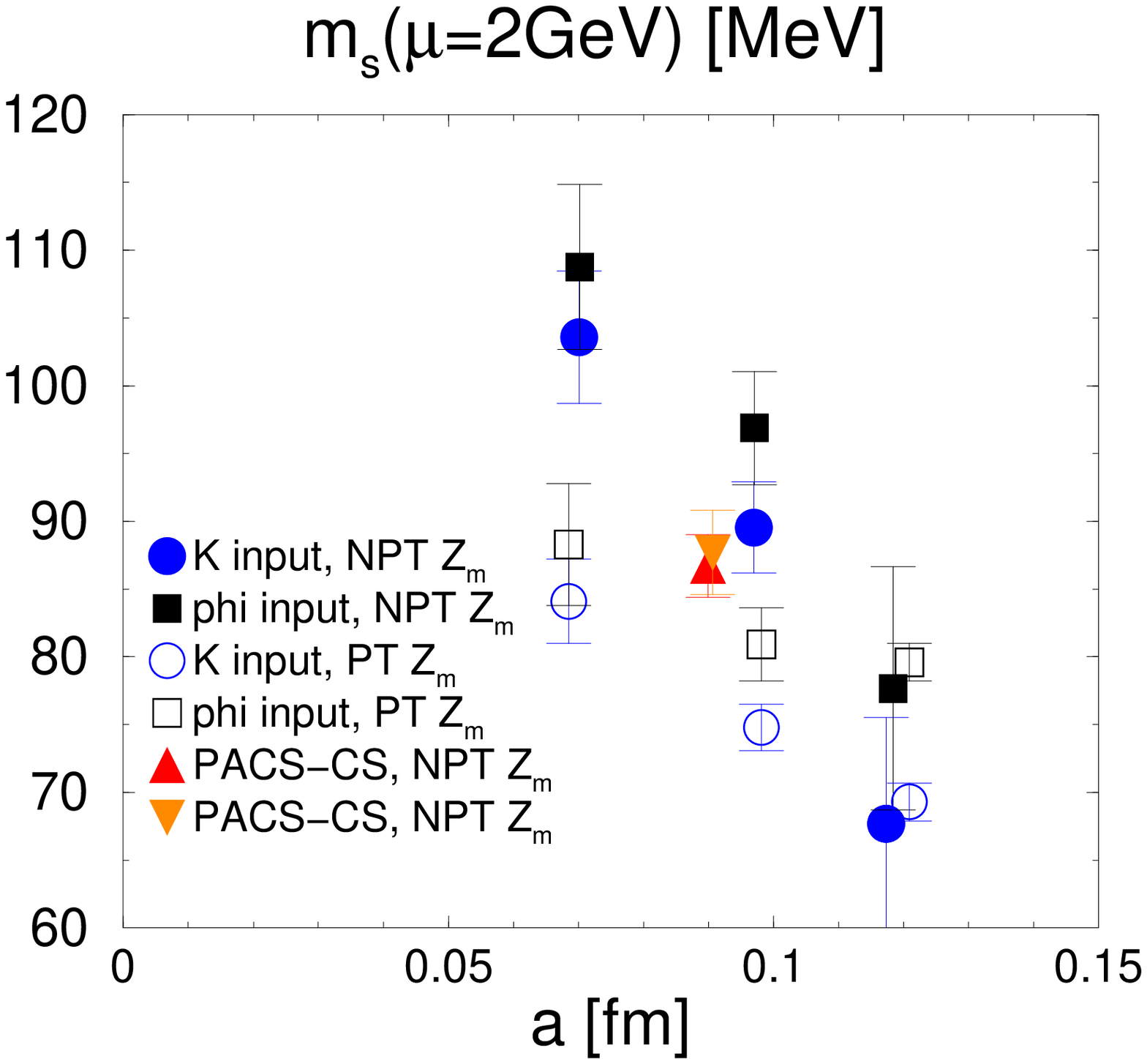}
 \caption{Scaling behavior of $m_{ud}^{\ovl{\rm MS}}$ (left) and
 $m_{s}^{\ovl{\rm MS}}$ (right).}
\label{fig:mud}
\end{center}
\end{figure}

As the last step we apply our renormalization factor to the bare PCAC
masses in Refs.~\cite{Ishikawa:2007nn,Aoki:2008sm,Kuramashi} to obtain
values for the renormalized quark masses.
We notice that the chiral perturbation theory was used to extract the
physical quark mass in Ref.~\cite{Ishikawa:2007nn,Aoki:2008sm}, while
the reweighting technique was adopted in Ref.~\cite{Kuramashi} to
evaluate the quark mass directly on the physical point.
The numerical results are given in Table \ref{tab:qmass}
and are plotted in Fig.~\ref{fig:mud} both for the averaged up and down
quark mass (left panel) and for the strange quark mass (right panel).
For the old CP-PACS/JLQCD work of Ref.~\cite{Ishikawa:2007nn}
results with $K$ and $\phi$ meson input are plotted (filled squares and
circles) together with perturbatively renormalized masses using
tadpole improved renormalization factor (open squares and circles).
The upward triangle represents the result for the more recent work of
PACS-CS evaluated directly on the physical point using the reweighting
technique \cite{Kuramashi}.
The downward triangle represent a result of chiral extrapolation from
the simulation point reaching down to $m_\pi=155$ MeV
\cite{Aoki:2008sm}.
\begin{table}[htb]
\begin{center}
\begin{tabular}{|c|c|c|c|c|}
\hline
$\beta$
 & $M_{ud}^{\rm RGI}$ & $m_{ud}^{\ovl{\rm MS}}$
 & $M_{s}^{\rm RGI}$  & $m_s^{\ovl{\rm MS}}$ \cr
\hline
$1.83$ \cite{Ishikawa:2007nn}
 & $3.30(38)$
 & $2.59(30)$
 & $86(10)$
 & $67.7(7.9)$
\cr
$1.90$ \cite{Ishikawa:2007nn}
 & $4.47(17)$
 & $3.47(13)$
 & $115.3(4.4)$
 & $89.5(3.4)$
\cr
$2.05$ \cite{Ishikawa:2007nn}
 & $5.29(24)$
 & $4.02(18)$
 & $136.1(6.5)$
 & $103.6(4.9)$
\cr
\hline
$1.90$ \cite{Aoki:2008sm}
 & $3.84(16)$ 
 & $3.05(12)$ 
 & $110.5(4.0)$
 & $87.7(3.1)$ 
\cr
$1.90$ \cite{Kuramashi}
 & $3.49(34)$
 & $2.78(27)$
 & $109.0(3.0)$
 & $86.7(2.3)$
\cr
\hline
\end{tabular}
\caption{Non-perturbatively renormalized mass for the averaged up and
 down quark and for the strange quark.
 $M^{\rm RGI}$ is the RGI mass and $m^{\ovl{\rm MS}}$ is that in the
 $\ovl{\rm MS}$ scheme at a scale $\mu=2$ GeV.
 The unit is in MeV.
 $K$-meson mass is used for physical input.}
\label{tab:qmass}
\end{center}
\end{table}

It is disappointing that the old CP-PACS/JLQCD results do not exhibit a
better scaling behavior by going from perturbative to non-perturbative
renormalization factor.
However, we should note a significant change in the average up and down
quark mass with the recent PACS-CS work (filled triangles).
This represents a systematic error due to chiral extrapolation of the
old CP-PACS/JLQCD work whose pion mass reached only $m_\pi\sim500$ MeV.
We should also note that the renormalization factor $Z_A$ involves a
large uncertainty at $\beta=1.83$ which is not reflected in the error
bar of Fig.~\ref{fig:mud}.
We feel that results at $\beta=2.05$ using simulation with physical pion
mass are needed to find the convincing values for light quark masses with
our approach.

\reseteqnum
\section{Conclusion}

We have presented a calculation of the quark mass renormalization factor
for the $N_f=2+1$ QCD in the mass independent Schr\"odinger functional
scheme in the chiral limit.
We calculate the SSF of the running mass on the lattice and applied a
perturbative improvement.
We find that the step scaling function shows a good scaling behavior and
the continuum limit may be taken safely with a constant extrapolation.
The non-perturbative SSF turned out to be almost consistent with the
perturbative two loops result.
We then derive the renormalization factor of the pseudo scalar density
at low energy and that of axial vector current.
Multiplying these factors we evaluate the non-perturbative
renormalization factors for the RGI mass and that in the $\ovl{\rm MS}$
scheme at a scale $\mu=2$ GeV.

Applying our non-perturbative renormalization factor to the present
PACS-CS simulation result evaluated directly at the physical point
yields
$m_{ud}^{\ovl{\rm MS}}(\mu=2\ {\rm GeV})=2.78(27)$ MeV for the average
up and down quark, and
$m_{s}^{\ovl{\rm MS}}(\mu=2\ {\rm GeV})=86.7(2.3)$ MeV for the strange
quark.
Simulations at weaker couplings under way will tell if these values stay
toward the continuum limit.

This work is supported in part by Grants-in-Aid of the Ministry of
Education (Nos.
10143538, 20105001, 20105002, 20105003, 20105005, 20340047,
20540248, 20740139, 21340049, 22105501, 22244018, 22740138).

\end{document}